\documentclass[12pt]{article}

\usepackage[cp866]{inputenc}
\usepackage{hyperref,amsfonts,amssymb,theorem}

\def\defeq {\stackrel{\mbox{\rm\small def}}{=}}

\def\bq{ \begin{equation} }
\def\eq{ \end{equation} }
\def\ben{ \begin{eqnarray} }
\def\en{ \end{eqnarray} }

\theorembodyfont{\sl}

\theorembodyfont{\rm}

\def\dsize{\displaystyle}

\def\p{ \partial }
\def\defeq {\stackrel{\mbox{\rm\small def}}{=}}

\def\bq{ \begin{equation} }
\def\eq{ \end{equation} }
\def\ben{ \begin{eqnarray} }
\def\en{ \end{eqnarray} }

\def\frac#1#2{{#1\over #2}}

\def\on#1#2{\mathop{\vbox{\ialign{##\crcr\noalign{\kern2pt}
$\scriptstyle{#2}$\crcr\noalign{\kern2pt\nointerlineskip}
\kern-2pt$\hfil\displaystyle{#1}\hfil$\crcr}}}\limits}

\newcommand{\be}{\begin{equation}}
\newcommand{\ee}{\end{equation}}

\newcommand{\bear}{\begin{array}}
\newcommand{\eear}{\end{array}}

\def\ba{\begin{array}}
\def\ea{\end{array}}

\title{Separation of variables on a non-hyperelliptic curve
}
\author{V.G. Marikhin and V.V. Sokolov}
\begin{document}
\maketitle

\section{Introduction}
\setcounter{equation}{0}

In the paper we consider several models that admit a separation of variables on
the following algebraic curve of genus 4:
\be\label{curve}
\Phi(\xi,Y)=s_6Y^6+l(\xi)Y^4+k(\xi)Y^2-S(\xi)=0,
\ee
where
\be\label{polyn}
s_6=\frac{\delta^6}{6!}S^{VI}(\xi),\qquad
k(\xi)=\frac{\delta^2}{10}S''(\xi)+4(\alpha \xi^{2}+e_2\xi+e_1),
\qquad l(\xi)= \frac{\delta^4}{4!}S^{IV}(\xi)-\frac{\delta^2}{2}k''(\xi).
\ee
Here $\alpha, \delta$ are parameters, $S$ is arbitrary sixth degree
polynomial. If $\delta=0$, then the curve is hyperelliptic of genus 3.

The class of curves (\ref{curve}) with $\delta\ne 0$ can be described as follows. Let
$P(\eta, \xi)=0$ be an arbitrary cubic. In the generic case it is an elliptic curve.
Then (\ref{curve}) is a double cover over the cubic defined by
the formula $\eta=\xi^{2}-\delta^{2} Y^{2}$.

To present out approach, we start with the Hamiltonian
\be \label{hamnew}
H=a p_{1}^{2}+ c p_{2}^{2}+ d  p_{1}+ e p_{2}+ f,
\ee
where
$$
a=-\frac{4 s_2 S(s_1)}{s_1-s_2},  \qquad c=\frac{4s_1 S(s_2)}{s_1-s_2},
\qquad
d=-s_1\frac{J}{s_1-s_2},\qquad e=-s_2\frac{J}{s_1-s_2},
$$
\be\label{fman}
\begin{array}{c}
\displaystyle{f=\frac{\delta^2}{40}\frac{s_{2}S''(s_{1})-s_{1}S''(s_{2})}{s_{1}-s_{2}}-
\frac{\delta^2}{4}\frac{s_{2}S'(s_{1})+s_{1}S'(s_{2})}{(s_{1}-s_{2})^2}+}\\[5mm]
\displaystyle{\frac{3\delta^2}{4}\frac{s_{2}S(s_{1})-s_{1}S(s_{2})}{(s_{1}-s_{2})^3}+
\alpha s_{1} s_{2}},
\end{array}
\ee
and
$$
J=2\delta\frac{\sqrt{S(s_1)}\sqrt{S(s_2)}}{s_1-s_2}
$$
as the most general model. It can be easily verified that $H$ commutes with the function
\be \label{intnew}
K=A  p_{1}^{2}+C p_{2}^{2}+ D p_{1}+ E p_{2}+ F,
\ee
where
$$
A=\frac{4S(s_1)}{s_1-s_2}, \qquad  C=-\frac{4S(s_2)}{s_1-s_2}, \qquad
D=\frac{J}{s_1-s_2},\qquad  E=\frac{J}{s_1-s_2},
$$
\be\label{VVmanint}
F=\frac{\delta^2}{40}\frac{S''(s_{2})-S''(s_{1})}{s_{1}-s_{2}}+
\frac{\delta^2}{4}\frac{S'(s_{1})+S'(s_{2})}{(s_{1}-s_{2})^2}
+\frac{3\delta^2}{4}\frac{S(s_{2})-S(s_{1})}{(s_{1}-s_{2})^3}-\alpha (s_{1}+s_{2})
\ee
with respect to the standard Poisson bracket $\lbrace p_{\alpha},s_{\beta}\rbrace=\delta_{\alpha
\beta}$.
If $\delta=0,$ then the terms in  $H$ and $K$ linear in momenta disappear and
$H$ belongs to the St\"ackel class. In this case
$s_{1},s_{2}$ are separated variables. The corresponding Abel's transformation on the
hyperelliptic curve of genus 3 has the form
$$
\frac{ds_1}{\sqrt{T(s_1)}}+\frac{ds_2}{\sqrt{T(s_2)}}=dt,\qquad
\frac{s_1ds_1}{\sqrt{T(s_1)}}+\frac{s_2ds_2}{\sqrt{T(s_2)}}=0.
$$
Here
$$
T(s)=16\,S(s) \,(\alpha s^2+ e_2 s+ e_1),
$$
and the constants $e_{1}$ and $e_{2}$ are values of the integrals $H$ and $K,$
respectively.

The main result of the paper is an explicit formula for the action in the case $\delta\ne 0$
depending on two parameters. We find it directly from the Hamilton-Jacobi equation.

In Sections 4-6 we find the action and a separation of variables for the
Kowalewski hyrostat, the Clebsch and the $so(4)$ Schottky-Manakov spinning
tops in a way analogous to that used in Sections 2-3 for the Hamiltonian
(\ref{hamnew}). Many significant papers (see, for instance, \cite{shot, clebsch, koet, man,
fedor, adler, vesel1, reysem,  misc, miscfom, wint, perel, komtsig} and references there-in)
are devoted to these models. In particular, it is known  \cite{bobenko, komtsig})
that the models are linked via some changes of variables. However according
to a common opinion of experts, a "satisfactory" separation of variables for
the above models is still not found.

For the Clebsch and the $so(4)$ Schottky-Manakov tops $S$ is a polynomial
of degree 4. An appearance of the corresponding algebraic curve of genus 3
could be predicted: the characteristic curve for the $3\times 3$-Lax
operator found by A. Perelomov \cite{perel} for the Clebsch top has the same
form. However in this paper we don't use any Lax representations and the
curve arises during computations in a rather natural manner.

The case $S = \hbox{const}$ also is of interest. The inhomogeneous hydrodynamic type
system (\ref{hydro}) corresponding to the pair (\ref{hamnew}),
(\ref{intnew}) coincides with the Gibbons-Tsarev equation  \cite{gibts}.
The separation of variables found in this paper gives rise to a
family of elliptic solutions for the Gibbons-Tsarev equation (see Section 7).

{\bf Acknowledgments.}  The authors are grateful to A.V.\ Borisov, Yu. N. Fedorov,
E.V. Ferapontov,  V.B.\ Kuznetsov, and F.\ Magri for
useful discussions.  The second author (V.S.) is grateful to the Max Planck
Institute for hospitality and financial support. The research was also partially supported
by RFBR grants 02-01-00431,  and NSh 1716.2003.1.

\section{Equations of motion}
\setcounter{equation}{0}

The Hamilton equations corresponding to (\ref{hamnew}) and (\ref{intnew}) have the form
\be\label{ham1}
\frac{d s_{1}}{d t}=2  a  p_{1}+
d, \qquad  \frac{d s_{2}}{d t}= 2  c  p_{2}+ e.
\ee
and
\be\label{ham2}
\frac{d s_{1}}{d \tau}=2  A  p_{1}+
D, \qquad  \frac{d s_{2}}{d \tau}= 2  C  p_{2}+ E.
\ee
Finding $p_{1},  p_{2}$ from (\ref{ham1}) and substituting them in the relations
$H=e_{1}$ and $K=e_{2},$ we get
\be\label{s12g1}
(s_1-s_2)\lbrack s_1\frac{\dot{s_1}^2}{S(s_1)}-s_2\frac{\dot{s_2}^2}{S(s_2)}\rbrack
+16(e_1-f)s_1s_2+\frac{4\delta^2}{(s_1-s_2)^3}(s_2^3\,S(s_1)-s_1^3\,S(s_2))=0
\ee
and
\be\label{s12g2}
(s_1-s_2)s_1s_2\left(\frac{\dot{s_1}^2}{S(s_1)}-\frac{\dot{s_2}^2}{S(s_2)}\right)
-4\delta\sqrt{S(s_1)S(s_2)}\left(
s_1^2\frac{\dot{s_1}}{S(s_1)}+s_2^2\frac{\dot{s_2}}{S(s_2)}\right)+
\ee
$$
16s_1s_2((e_1-f)(s_1+s_2)+(e_2-F)s_1s_2)-
\frac{4\delta^2}{(s_1-s_2)^3}\left(s_2^3\,S(s_1)(s_1-2s_2)-s_1^3\,S(s_2)(s_2-2s_1)\right)=0,
$$
where $f$ and $F$ are defined by (\ref{fman}), (\ref{VVmanint}).

One of the basic technical problems is a rewriting of this system in a
compact form. The straightforward elimination say $\dot{s_2}$ leads to a
cumbersome equation of degree 4 with respect to $\dot{s_1}$.

{\bf Proposition 1.} {\it Let} $(u,v)$ {\it be any solution of the following system of equations:}
\be\label{uvsys}
L(s_2)u^2-2L(s_1)v-M(s_2,s_1)=0,\qquad  L(s_1)v^2-2L(s_2)u-M(s_1,s_2)=0,
\ee
{\it where}
$$
M(x,y)=3L(y)+L'(y)(x-y)+k(y)(x-y)^2,\qquad L(x)=\delta^2S(x).
$$
{\it Then the derivatives}
\be\label{uv}
\dot s_{1}=-J\frac{s_2\,u+s_1}{s_1-s_2},\qquad
\dot s_{2}=-J\frac{s_1\,v+s_2}{s_1-s_2}
\ee
{\it satisfy the system} (\ref{s12g1}), (\ref{s12g2}).

The elimination of variables $p_{1}, p_{2}$ from (\ref{ham1}), (\ref{ham2}) leads to the
inhomogeneous hydrodynamic type system
\be\label{hydro}
(s_{1})_{t}+s_2 (s_{1})_{\tau}=-J,\qquad
(s_{2})_{t}+s_1 (s_{2})_{\tau}=J.
\ee
It follows from this formula and from (\ref{uv}) that
\be\label{symmet}
(s_{1})_{\tau}=J\frac{u+1}{s_1-s_2},\qquad
(s_{2})_{\tau}=J\frac{v+1}{s_1-s_2}.
\ee
Moreover, we find from (\ref{uv}) that
\be\label{pi12}
p_1=\frac{J\,u}{8S(s_1)},
\qquad p_2= -\frac{J\,v}{8S(s_2)}.
\ee

The following parameterization for solutions of (\ref{uvsys}) turns out to be crucial for
the separation of variables.

{\bf Proposition 2.} {\it Let} $z_{1}, z_{2}, z_{3}$  {\it be solutions of the equation}
\be\label{eZ}
z^6L(s_2)^2-L(s_2)M(s_2,s_1)z^4+M(s_1,s_2)L(s_1)z^2-L(s_1)^2=0
\ee
{\it such that}
$$
\frac{L(x)}{L(y)}=z_1z_2z_3,\qquad \frac{M(y,x)}{L(y)}=z_1^2+z_2^2+z_3^2,\qquad
\frac{M(x,y)}{L(x)}=\frac{1}{z_1^2}+\frac{1}{z_2^2}+\frac{1}{z_3^2}.
$$
{\it Then}
\be\label{uvsol}
\begin{array}{c}
\displaystyle{(u_1,v_1)=(z_1+z_2+z_3,\,\frac{1}{z_1}+\frac{1}{z_2}+\frac{1}{z_3}),\qquad
(u_2,v_2)=(z_1-z_2-z_3,\,\frac{1}{z_1}-\frac{1}{z_2}-\frac{1}{z_3})},\\[5mm]
\displaystyle{ (u_3,v_3)=(z_2-z_1-z_3,\,\frac{1}{z_2}-\frac{1}{z_1}-\frac{1}{z_3}),\qquad
(u_4,v_4)=(z_3-z_1-z_2,\, \frac{1}{z_3}-\frac{1}{z_1}-\frac{1}{z_2})}
\end{array}
\ee
{\it are solutions of} (\ref{uvsys}).

\section{Separation of variables.}
\setcounter{equation}{0}
To separate the variables in the case $\delta\ne 0$, we find the action
function $\tilde S(s_{1},s_{2})$ in an explicit form.

Let us consider the system of equations
$$
\Phi(\xi,Y)=0,\qquad
Y^2=\frac{1}{\delta^2}(\xi-s_1)(\xi-s_2),
$$
where the polynomial $\Phi(\xi,Y)$ is given by (\ref{curve}). It is easy to
verify that if we substitute to the first equation the expression for $Y^{2}$ taken from the
second equation, then leading powers of $\xi$ cancel and the equation for $\xi$
turns out to be cubic. By $\, \xi_{i}(s_{1},s_{2}), \quad i=1,2,3\,$ denote the roots
of this equation.

{\bf Theorem.} {\it The function}
\be\label{fullsep}
\tilde{S}(s_{1},s_{2},e_{1},e_{2})=\frac{1}{4}\sum\limits_{n=1}^3\left[\delta\, \hbox{arctanh}\,
\frac{\xi_n-\frac{1}{2}(s_1+s_2)}{\delta \,Y(\xi_n)}
-\int\limits^{\xi_n}\frac{d\xi}{Y(\xi)}\right]
\ee
{\it satisfies the Hamilton-Jacobi equation}
$$
H\left(\frac{\p \tilde{S}}{\p s_{1}},\frac{\p \tilde{S}}{\p s_{2}}, s_{1},
s_{2}\right)=e_{1}, \qquad K\left(\frac{\p \tilde{S}}{\p s_{1}},\frac{\p \tilde{S}}{\p s_{2}}, s_{1},
s_{2}\right)=e_{2},
$$
{\it where the functions} $H(p_{1},p_{2},s_{1},s_{2})$ and  $K( p_{1}, p_{2},s_{1},s_{2})$
{\it are defined by formulas} (\ref{hamnew})-(\ref{intnew}).

{\bf Proof.} It follows from (\ref{pi12}) that the partial derivatives of the action
$\tilde{S}$ are given by:
\be\label{pi13}
\frac{\partial \tilde{S}}{\partial s_1}=
\frac{J\,u}{8S(s_1)},
\qquad
\frac{\partial \tilde{S}}{\partial s_2}=
-\frac{J\,v}{8S(s_2)}.
\ee
The compatibility condition for this system has the form
$$S(s_2)\frac{\partial}{\partial s_2}(Ju)+S(s_1)\frac{\partial}{\partial s_1}(Jv)=0.$$
The latter relation follows from the identities
\be\label{togRL}
\begin{array}{c}
\left(L'(y)+M_x(x,y)\right)(x-y)+6L(y)-2M(x,y)=0,\\[3mm]
(x-y)\left(M_y(x,y)+M_x(y,x)\right)+4(M(x,y)-M(y,x))=0,
\end{array}
\ee
which can be verified straightforwardly.

Consider an auxiliary Hamiltonian of the form (compare with formula (\ref{eZ}))
\be\label{Hps}
\tilde{H}=p_1^3L(s_{1})+p_1^2p_2M(s_{2},s_{1})+p_1p_2^2M(s_{1},s_{2})+p_2^3L(s_{2}).
\ee

It can easily be checked that identities (\ref{togRL}) are fulfilled if the condition  $\lbrace
\tilde{H},\tilde{K}\rbrace=0$ holds, where
$$
 \tilde{K}=-(s_{1}-s_{2})^2p_1p_2.
$$

It is easy to verify that if $\lbrace p_i,s_j\rbrace=\delta_{ij}$, then the Poisson brackets between
functions
\be\label{axp} a_0=p_1+p_2,\qquad
a_1=s_1p_1+s_2p_2,\qquad a_2=s_1^2p_1+s_2^2p_2 \ee
are given by
\be\label{sl2}
\lbrace a_0,a_1\rbrace=a_0,\qquad \lbrace a_0,a_2\rbrace=2a_1,\qquad \lbrace
a_1,a_2\rbrace=a_2. \ee
The expression $Q=a_1^2-a_0 a_2$ is a Casimir function for the linear
$sl(2)$-brackets (\ref{sl2}). It turns out that $\tilde{K}$ rewritten in variables (\ref{axp})
coincides with $Q$. The Hamiltonian $\tilde{H}$ can also be expressed in terms of variables (\ref{axp})
only:
$$
\tilde{H}=a_0^3L\left(\frac{a_1}{a_0}\right)-\delta^2(\alpha a_2+e_2a_1+e_1a_0)
+\frac{\delta^4}{4}\frac{a_1^2}{a_0}.
$$
It is seen from this formula that the variable
$$\xi=\frac{a_1}{a_0}=\frac{p_1s_1+p_2s_2}{p_1+p_2}$$
should play a key role in a description of properties of polynomials (\ref{eZ}), (\ref{Hps}).
Suppose $\tilde{H}(p_{1},p_{2})=0;$ then a solution $z$ for equation (\ref{eZ}) can be expressed
in terms of $\xi$ by the formula
\be\label{zzz}
z=\sqrt{\frac{s_2-\xi}{s_1-\xi}}\sqrt{\frac{S(s_1)}{S(s_2)}}.
\ee
Equation (\ref{eZ}) is cubic with respect to $\xi$. It can be rewritten as (\ref{curve}),
where
\be\label{yY}
Y^2\defeq\frac{1}{\delta^2}(\xi-s_1)(\xi-s_2).
\ee

Our task is to find the solution of (\ref{pi13}) explicitly. According to formulas
(\ref{uvsol}), it suffices for that to solve the system of equation
\be\label{sysz}
\begin{array}{c}
\displaystyle{\frac{\partial \sigma}{\partial s_1}= \frac{J}{8S(s_1)}z=
\frac{\delta}{4(s_1-s_2)}\sqrt{\frac{s_2-\xi}{s_1-\xi}}},\\[5mm]
\displaystyle{ \frac{\partial  \sigma}{\partial s_2}=
-\frac{J}{8S(s_2)}\frac{1}{z}=-\frac{\delta}{4(s_1-s_2)}\sqrt{\frac{s_1-\xi}{s_2-\xi}}}.
\end{array}
\ee
Here $z(s_{1},s_{2})$ is arbitrary root of equation (\ref{eZ}) and $\xi$ is the
corresponding (see (\ref{zzz})) value of $\xi(s_{1},s_{2})$. The
action function $\tilde S$ required can be obtained as the sum of the three solutions
for (\ref{sysz}), corresponding to the three branches of the function $\xi(s_{1},s_{2})$.

First, we find a function $\sigma_0$ such that (\ref{sysz}) is fulfilled under condition
that $\xi$ is a parameter, which does not depend on $s_1,s_2.$ It is not hard to see that
$$\sigma_0(\xi,s_1,s_2)=\frac{1}{4}\delta\, \hbox{arctanh}\,
\frac{\xi-\frac{1}{2}(s_1+s_2)}{\sqrt{(\xi-s_1)(\xi-s_2)}}$$

Taking into account the formula
$$\frac{\partial}{\partial\xi}\sigma_0=\frac{1}{4}\frac{1}{\sqrt{(\xi-s_1)(\xi-s_2)}}$$
and replacing  $\sqrt{(\xi-s_1)(\xi-s_2)}$ by $Y\delta,$ we obtain expression (\ref{fullsep}) for the
action function.

Differentiating the action with respect to the parameters and taking into account
that
$$\frac{\partial\tilde{S}}{\partial e_1}=t+c_{1},\qquad \frac{\partial\tilde{S}}{\partial e_2}=c_{2},
$$
we finally get
\be\label{abelman}
dt=\sum\limits_{n=1}^3\omega_1(\xi_n),\qquad
0=\sum\limits_{n=1}^3\omega_2(\xi_n).
\ee
Here $\omega_1,\;\omega_2$ belong to a basis
\be\label{om1234}
\omega_1(\xi)=\frac{d\xi}{Z},\qquad \omega_2(\xi)=\frac{\xi\,d\xi}{Z},\qquad
\omega_3(\xi)=\frac{(\xi^2-\delta^2 Y^2)d\xi}{Z},\qquad
\omega_4(\xi)=\frac{Y\,d\xi}{Z}
\ee
of holomorphic differentials on the curve $\Phi(\xi,Y)=0$. In (\ref{om1234}) we use the notation
$Z\defeq\frac{\p \Phi}{\p Y}$.

The differential $\omega_{4}$ plays a special role.
In variables  $(\xi,\eta)$, where
\be\label{eta}
\eta=\xi^2-\delta^2 Y^2,
\ee
the curve (\ref{curve}) becomes the following cubic
\be\label{qubic}
s_6\eta^3+s_5\eta^2\xi+\frac{1}{5}s_4\eta(\eta+4\xi^2)+\frac{1}{5}s_3\xi(\eta+2\xi^2)+
\frac{1}{5}s_2(\eta+4\xi^2)+s_1\xi+s_0+\frac{4}{\delta^2}(\eta-\xi^2)(\alpha\eta+e_2\xi+e_1)=0.
\ee
The holomorphic differential of the elliptic curve (\ref{qubic}) coincides
with $\omega_{4}$ up to transformation (\ref{eta}).

It follows from (\ref{yY}) that the functions  $Y(\xi_{i})$ are linked by the relation
\be\label{Y123}
\frac{1}{\delta^2}=\frac{Y^{2}(\xi_1)}{(\xi_1-\xi_2)(\xi_1-\xi_3)}+
\frac{Y^{2}(\xi_2)}{(\xi_2-\xi_3)(\xi_2-\xi_1)}+\frac{Y^{2}(\xi_3)}{(\xi_3-\xi_1)(\xi_3-\xi_2)},
\ee
which allows us to determine the two functions $s_{1}(t),  s_{2}(t)$ starting from
$\xi_{1}(t),  \xi_{2}(t),  \xi_{3}(t).$  Thus  we have three conditions
(\ref{abelman}), (\ref{Y123}) to determine three functions
$\xi_{i}(t)$.

Condition (\ref{Y123}) can be rewritten in the variables $(\xi,\eta)$ as
$$\eta_1(\xi_2-\xi_3)+\eta_2(\xi_3-\xi_1)+\eta_3(\xi_1-\xi_2)=0.$$
This means that the corresponding points $(\xi_i,\eta_i)$ belongs to the intersection of
elliptic curve (\ref{qubic}) and the straight line $\eta=\xi(s_1+s_2)-s_1s_2.$
\hspace{0.5cm}
\section{Schottky-Manakov spinning
top.}
\setcounter{equation}{0}

It is well-known (see, for example \cite{vesel1, misc}) that the Hamiltonian
$$
H=(\vec{S_1},A\vec{S_1})+2(\vec{S_1},B\vec{S_2})+(\vec{S_2},A\vec{S_2}),
$$
where $A=\hbox{diag}(a_1,a_2,a_3),\;B=\hbox{diag}(b_1,b_2,b_3)$, commutes with a quadratic polynomial
$K$ with respect to the spin Poisson brackets
\be\label{puas}
 \lbrace S_i^{\alpha},S_j^{\beta}\rbrace=
\kappa\,\varepsilon_{\alpha\beta\gamma}S_i^{\gamma}\delta_{ij} \ee
iff
\be
b_1^2(a_2-a_3)+b_2^2(a_3-a_1)+b_3^2(a_1-a_2)+(a_1-a_2)(a_2-a_3)(a_3-a_1)
=0.
\ee
Here $\varepsilon_{\alpha\beta\gamma}$ is the totally skew-symmetric tensor, $\kappa$ is a parameter.

Since $H$ and $K$ may be replaced by arbitrary linear combinations of
$H$, $K$ and the Casimir functions
$$
J_{1}=(\vec{S}_1,\vec{S}_1), \qquad J_{2}=(\vec{S}_2,\vec{S}_2)
$$
for the brackets (\ref{puas}), the integral $K$ can be reduced  \cite{wint} to the form
$$
K=2(\vec{S_1},\hat{C}\vec{S_2}),\qquad
C=\hbox{diag}(\alpha_1,\alpha_2,\alpha_3).
$$
Without loss of generality the matrices $A$ and $B$ defined the Hamiltonian $H$
can be choosen as follows
$$A=\hbox{diag}(-\alpha_1^2, -\alpha_2^2,-\alpha_3^2),\qquad
B=\hbox{diag}(\alpha_2\alpha_3+\lambda\alpha_1, \,\,
\alpha_3\alpha_1+\lambda\alpha_2,\,\, \alpha_1\alpha_2+\lambda\alpha_3),$$
The arbitrary parameter $\lambda$ corresponds to the shift of $H$ by $\lambda\, K.$

{\bf Reduction to the standard brackets.}
Let us fix the values of the Casimir functions: $(\vec{S}_k,\vec{S}_k)=j_k^2.$ Then
the formulas
\be\label{spar}
\vec{S}_k=p_k\vec{K}(q_k)+\frac{j_k}{2}\vec{K}'(q_k),\qquad
\hbox{\rm where} \qquad \vec{K}(q)=((q^2-1),\,i(q^2+1),\,2q),\; \ee
define a transformation of the Poisson manifold with coordinates $\vec{S}_1,\vec{S}_2$
and brackets (\ref{puas}), where $\kappa=-2 i$,
to the manifold with coordinates $p_{1},p_{2}, q_{1},q_{2}$ and canonical Poisson
brackets $\lbrace p_{\alpha},q_{\beta}\rbrace=\delta_{\alpha \beta}.$ As a result of this
transformation we get
\be
\begin{array}{c}
\displaystyle{ H=p_1^2r(q_1)+\frac{j_1}{2}p_1\,r'(q_1)+\frac{j_1^2}{12}r''(q_1)+
p_2^2r(q_2)+\frac{j_2}{2}p_2\,r'(q_2)+\frac{j_2^2}{12}r''(q_2)}+\\[3mm]
\displaystyle{ 2\left(p_1+\frac{j_1}{2}\frac{\partial}{\partial q_1}\right)
\left(p_2+\frac{j_2}{2}\frac{\partial}{\partial q_2}\right)Z(q_1,q_2)},
\end{array}  \label{hpq}
\ee
\be\label{kpq} K=2\left(p_1+\frac{j_1}{2}\frac{\partial}{\partial q_1}\right)
\left(p_2+\frac{j_2}{2}\frac{\partial}{\partial q_2}\right)W(q_1,q_2),\ee
where
$$
r(x)\defeq (\vec{K}(x),\,A\vec{K}(x))=-\alpha_1^2(x^2-1)^2+\alpha_2(x^2+1)^2-4\alpha_3x^2,
$$
\be \label{zz}
\begin{array}{c} Z(x,y)\defeq (\vec{K}(x),\,B\vec{K}(y))=\\
(\alpha_2\alpha_3+\lambda\alpha_1)(x^2-1)(y^2-1)-(\alpha_3\alpha_1+\lambda\alpha_2)(x^2+1)(y^2+1)
+4(\alpha_1\alpha_2+\lambda\alpha_3)xy,
\end{array}
\ee
\be \label{ww}W(x,y)\defeq (\vec{K}(x),\,C\vec{K}(y))=
\alpha_1(x^2-1)(y^2-1)-\alpha_2(x^2+1)(y^2+1)+4\alpha_3xy.
\ee
It is possible to check that
\be \label{barw}
Z^2(x,y)-r(x) r(y)=W(x,y) \bar W(x,y),
\ee
where $\bar W$ is a polynomial quadratic in each of variables.

{\bf Remark. } Note that the more general Hamiltonian
$$
H=\sum\limits_{ij,\mu,\nu}S_i^{\mu}S_j^{\nu}c^{ij}_{\mu\nu},\qquad i,j=1,...,N,\qquad
\mu,\nu=x,y,z,
$$
describing an interaction of $N$ spins, can be reduced to
$$
H=\sum\limits_{ij} g^{ij}p_ip_j+\sum\limits_{i} a^ip_i+v,
$$
where
$$
\quad g^{ij}=\sum\limits_{\mu,\nu}K^{\mu}(x^i)K^{\nu}(x^j)c^{ij}_{\mu\nu},
$$
$$a^i=\sum\limits_k j_k\frac{\partial g^{ik}}{\partial x^k}-\frac{1}{2}j_i\frac{\partial g^{ii}}{\partial
x^i},
$$
$$v=\frac{1}{4}\sum\limits_{ik}j_ij_k\frac{\partial^2 g^{ik}}{\partial x^i\partial x^k}-
\frac{1}{6}\sum\limits_kj_k^2\frac{\partial^2 g^{kk}}{\partial x^k\partial x^k}$$
by a similar transformation
$$\vec{S}_k=p_k\vec{K}(x^{k})+\frac{j_k}{2}\vec{K}'(x^{k}),  \qquad
k=1,...,N.
$$
In terms of the coordinates and the velocities the Lagrangian and the energy
read as
$$
L=\frac{1}{4}\sum\limits_{ij} g_{ij}(\dot{x}^i-a^i)(\dot{x}^j-a^j)-v,\qquad
E=\frac{1}{4}\sum\limits_{ij} g_{ij}(\dot{x}^i\dot{x}^j-a^ia^j)+v,\quad
$$
where $\sum\limits_jg_{ij}g^{jk}=\delta_i^k.$

An additional integral of the form
$$
K=\sum\limits_{ij}
G^{ij}p_ip_j+\sum\limits_{i} A^ip_i+V
$$
exists iff
$$
\sum\limits_{k} \frac{\p G^{ij}}{\p x^{k}} g^{kl}-\frac{\p g^{ij}}{\p x^{k}}G^{kl}=0,\qquad
\sum\limits_{k}a^k\frac{\p G^{ij}}{\p x^{k}}-2  \frac{\p a^{i}}{\p x^{k}}  G^{kj}-
A^k\frac{\p g^{ij}}{\p x^{k}}+2 \frac{\p A^{i}}{\p x^{k}} g^{kj}=0,\quad
$$
$$\sum\limits_{k} a^k \frac{\p A^{i}}{\p x^{k}}-A^k \frac{\p a^{i}}{\p x^{k}}+2g^{ik}
\frac{\p V}{\p x^{k}}-2G^{ik}\frac{\p v}{\p x^{k}}=0,
\qquad \sum\limits_{k} a^k \frac{\p V}{\p x^{k}}-A^k \frac{\p v}{\p x^{k}}=0.$$

{\bf Diagonalization of the quadratic part.}
Under transformation (\ref{spar}) the functions $H$ and $K$ take the form
\be \label{quadr}
H=a p_{1}^{2}+2 b p_{1} p_{2}+c
p_{2}^{2}+d p_{1}+e p_{2}+f,
\ee
\be \label{quadr1}
K=A p_{1}^{2}+2 B p_{1} p_{2}+C
p_{2}^{2}+D p_{1}+E p_{2}+F,
\ee
where the coefficients are some (in our case rational) functions of the variables
$q_{1},q_{2}$.

General formulas related to a pair of commuting Hamiltonians quadratic in
momenta and some examples can be found in
\cite{winter1,ferap,ferapfor,winter2}.

The class of Hamiltonians (\ref{quadr}) is invariant with respect to {\it
canoniacal} transformations of the form
$$
 p_{1}=k_{1} \hat p_{1}+k_{2} \hat p_{2}+k_{3}, \qquad p_{2}=\bar k_{1} \hat p_{1}+\bar k_{2} \hat
 p_{2}+\bar
k_{3}, \qquad \qquad q_{1}=\phi, \qquad q_{2}=\bar \phi,
$$
where $k_{i}, \bar k_{i}, \phi, \bar \phi$ are some functions of
$\hat q_{1},\hat q_{2}$. It is easily seen that the functions $k_{1},k_{2},\bar k_{1},\bar k_{2}$
can be uniquely expressed through $\phi, \bar \phi$ from the
condition $\lbrace p_{\alpha},q_{\beta}\rbrace=
\delta_{\alpha \beta}$:
$$
k_{1}=\frac{\p \bar \phi}{\p \hat q_{2}} W^{-1},\quad  k_{2}=-\frac{\p \bar \phi}{\p \hat q_{1}}
W^{-1}, \qquad
\bar k_{1}=-\frac{\p \phi}{\p \hat q_{2}} W^{-1},\quad  \bar k_{2}=\frac{\p \phi}{\p \hat
q_{1}},
$$
where
$$
W=\frac{\p \bar \phi}{\p \hat q_{2}} \frac{\p \phi}{\p \hat
q_{1}}-\frac{\p \bar \phi}{\p \hat q_{1}} \frac{\p \phi}{\p \hat
q_{2}}.
$$

The fact that the coefficient of $\hat p_{1} \hat p_{2}$ in the transformed
Hamiltonian (\ref{quadr}) vanishes is equivalent to
\be \label{cond1}
a(\phi, \bar \phi) \frac{\p \bar \phi}{\p \hat q_{1}}  \frac{\p \bar \phi}{\p \hat
q_{2}}+c(\phi, \bar \phi) \frac{\p \phi}{\p \hat q_{1}}  \frac{\p \phi}{\p \hat
q_{2}}=b(\phi, \bar \phi)\left(\frac{\p \bar \phi}{\p \hat q_{2}} \frac{\p \phi}{\p \hat q_{1}}+
\frac{\p \bar \phi}{\p \hat q_{1}} \frac{\p \phi}{\p \hat q_{2}} \right).
\ee
Analogously, the condition
\be \label{cond2}
A(\phi, \bar \phi) \frac{\p \bar \phi}{\p \hat q_{1}}  \frac{\p \bar \phi}{\p \hat
q_{2}}+C(\phi, \bar \phi) \frac{\p \phi}{\p \hat q_{1}}  \frac{\p \phi}{\p \hat
q_{2}}=B(\phi, \bar \phi)\left(\frac{\p \bar \phi}{\p \hat q_{2}} \frac{\p \phi}{\p \hat q_{1}}+
\frac{\p \bar \phi}{\p \hat q_{1}} \frac{\p \phi}{\p \hat q_{2}} \right)
\ee
guarantees that the coefficient of $\hat p_{1} \hat
p_{2}$ in the transformed integral (\ref{quadr1}) is equal to zero. The existence of
a solution $\phi, \bar \phi$ for system (\ref{cond1}), (\ref{cond2}) is obvious. Our
goal is to find the functions $\phi, \bar \phi$ for the pair (\ref{hpq}), (\ref{kpq}) explicitly.

Let us introduce the Kowalewski variables  $s_{1}(q_{1},q_{2})$ and $s_{2}(q_{1},q_{2})$
as the roots of the equation
$$
W(q_{1},q_{2})\, s^2+2 Z(q_{1},q_{2})\, s+\bar{W}(q_{1},q_{2})=0,
$$
where $W$ and $Z$ are given by (\ref{ww}),(\ref{zz}) and the polynomial $\bar W$
is defined by (\ref{barw}).

{\bf Proposition 3.} {\it
In the variables} $s_{1},$ $s_{2}$ {\it the functions} $H$ {\it and} $K$ {\it have the form}
(\ref{hamnew})-(\ref{intnew}),
{\it where polynomials} $S$ {\it and} $R$ {\it are defined by the formulas}
$$
S(x)=-(x+\lambda-\alpha_1-\alpha_2-\alpha_3)(x+\lambda+\alpha_1+\alpha_2-\alpha_3)
(x+\lambda+\alpha_1-\alpha_2+\alpha_3)(x+\lambda-\alpha_1+\alpha_2+\alpha_3)
$$
{\it and}
$$
\alpha=\frac{1}{4}(\frac{\delta^2}{5}-\nu^2),
\qquad \delta=j_{1}-j_{2}, \qquad\nu=j_1+j_2.
$$
The values $h$ and $k$ of the integrals $H$ and $K$ are related to the
constants $e_{1}, e_{2}$ from (\ref{polyn}) by
$$
h=e_1+\frac{1}{12}\alpha S''(0),\qquad
k=e_2+\frac{1}{12}\alpha S'''(0).
$$

Thus the general scheme from Section 3 can be applied to the Schottky-Manakov spinning
top.
\hspace{0.5cm}
\section{The Kowalewski gyrostat.}
\setcounter{equation}{0}
In this Section we use some formulas from \cite{komtsig} that describe a dynamical system
for the Kowalewski gyrostat in the Kowalewski variables.

The Hamiltonain structure for the gyrostat is defined by the $e(3)$-Poisson brackets
$$
 \bigl\{M_i\,,M_j\,\bigr\}=\varepsilon_{ijk}M_k\,, \qquad
\bigl\{M_i\,,\gamma_j\,\bigr\}=\varepsilon_{ijk}\gamma_k \,, \qquad
\bigl\{\gamma_i\,,\gamma_j\,\bigr\}=0,
$$
where $\varepsilon_{ijk}$ is the totally skew-symmetric tensor.
The brackets possess the Casimir functions
\begin{equation}\label{caz}
A=\sum_{k=1}^3 \gamma_k^2, \qquad
   B=\sum_{k=1}^3  \gamma_k M_k .
\end{equation}

The Hamiltonian for the gyrostat is as follows:
\be
 \label{Hgir}
H=\frac12(M_1^2+M_2^2+
2M_3^2-2\lambda M_3 )+c \gamma_1\,,\\
\ee
where $c$ and $\lambda$ are constants. An additional integral of motion is
given by
\be
 \label{Kgir}
K=\xi_{1} \xi_{2}+4\lambda\Bigl((M_3-\lambda)z_{1}
z_{2} -(z_{1} +z_{2} )c \gamma_3\Bigr),
\ee
where
$$
    \xi_{1}=z_{1}^2-2c(\gamma_1+i\gamma_2)\,, \qquad  \xi_{2}=z_{2} ^2-2c (\gamma_1-i\gamma_2)
$$
and
$$
z_1=M_1 + i M_2, \qquad z_2=M_1-i M_2.
$$
We put
$$ R(z_{1} ,z_{2} )=
 z_{1} ^2 z_{2} ^2-2h(z_{1} ^2+z_{2} ^2) -4c b(z_{1} +z_{2} )-4 c^2\,
a+k.
$$
Here $a$, $b$, $h$ and $k$ are values of integrals (\ref{caz}), (\ref{Hgir}), and (\ref{Kgir}).
The Kowalewski variables are defined as follows
$$
\label{s1,2} s_{1,2}=\frac{R(z_{1} ,z_{2} ) \pm\sqrt{R(z_{1}
,z_{1} )R(z_{2} ,z_{2} )}} {2(z_{1} -z_{2} )^2}.
$$
It turns out \cite{komtsig} that
\be \label{Hs} h=
\frac{s_1-s_2}2\,\left(\frac{\dot{s}_1^2}{\varphi_1}
-\frac{\dot{s}_2^2}{\varphi_2}\right)-\frac{s_1+s_2}2\,,
\ee
\be
\label{Ks}
\frac{k}4=(2h+s_1+s_2)\lambda^2-\lambda\sqrt{-\varphi_1\varphi_2}
\left(\dsize\frac{\dot{s}_1}{\dsize\varphi_1}+\frac{\dot{s}_2}{\varphi_2}\right)
+(s_1-s_2)\left(\frac{s_2\dot{s}_1^2}{\varphi_1}
-\frac{s_1\dot{s}_2^2}{\varphi_2}\right)-s_1s_2+h^2\,.
\ee
Here $\varphi_i=S(s_{i})$,
$$
S(s)=4s^3-8h\,s^2+ 4h^2\,s-k\,s+ 4c^2 a\,s+4c^2\,b.
$$

Substituting the expressions (\ref{symmet})
$$
\dot{s}_1=-\frac{1}{4}\frac{J}{s_1-s_2}(u+1)
,\qquad \dot{s}_2=-\frac{1}{4}\frac{J}{s_1-s_2}(v+1)
$$
in (\ref{Hs}), (\ref{Ks}) for the velocities, we get just
(\ref{uvsys}), where
$$
k(x)=4(x+h)^2+4\delta^2(x-2h)-k, \qquad \lambda=i\delta.
$$
Thus in this case the scheme for the separation of variables described in Section 3 is applicable.
The curve $\Phi(Y,\xi)=0$ is of genus 3. The differentials $\omega_1,\omega_2,\omega_4$ form a
basis of holomorphic differentials.

Notice that the substitution $\delta=-i\lambda$ in
(\ref{fullsep}) results the change $arctanh\rightarrow arctan$ and the
action remains to be real.
\hspace{0.5cm}
\section{The Clebsch spinning top.}
\setcounter{equation}{0}
The Clebsch spinning top is defined by the Hamiltonian
$$
H=\frac{1}{2}(J_1^2+J_2^2+J_3^2)+\frac{1}{2}(\lambda_1x_1^2+\lambda_2x_2^2+\lambda_3x_3^2)
$$
which commutes with respect to the  $e(3)$-Poisson brackets
$$
\lbrace J_i,J_j\rbrace=i\,\varepsilon_{ijk}J_k,\qquad \lbrace
x_i,x_j\rbrace=0,\qquad\lbrace J_i,x_j\rbrace=i\,\varepsilon_{ijk} x_k
$$
with the first integral
$$
K=(\lambda_1J_1^2+\lambda_2J_2^2+\lambda_3J_3^2)-
\lambda_1\lambda_2\lambda_3(\frac{x_1^2}{\lambda_1}+\frac{x_2^2}{\lambda_2}+\frac{x_3^2}{\lambda_3}).
$$

Let us fix the values of the Casimir functions as follows
$$x_1^2+x_2^2+x_3^2=a ^2,\qquad J_1x_1+J_2x_2+J_3x_3=l.$$

Using the parameterization
$$
J_1=\frac{1}{2}p_1(1-q_1^2)+\frac{1}{2}p_2(1-q_2^2)+\frac{l}{a }\,q_1,\;
J_2=\frac{i}{2}p_1(1+q_1^2)+\frac{i}{2}p_2(1+q_2^2)-i\frac{l}{a }\,q_1,\;
J_3=p_1q_1+p_2q_2-\frac{l}{a },
$$
$$
x_1=a \frac{1-q_1q_2}{q_1-q_2},\qquad
x_2=ia \frac{1+q_1q_2}{q_1-q_2},\qquad x_3=a \frac{q_1+q_2}{q_1-q_2},
$$
we express $H$ and $K$ in terms of canonically
conjugated variables $p_1,q_1,\;p_2,q_2$:
$$
H=-\frac{1}{2}(x-y)^2p_1p_2+\frac{l}{a }\,p_2(x-y)+\frac{a ^2}{2}\frac{W(x,y)}{(x-y)^2}
+\frac{1}{2}(\lambda a^2+\frac{l^2}{a ^2})
$$
$$
K=\frac{1}{4}(R(x)p_1^2+R(y)p_2^2+2\,p_1p_2W(x,y))-\frac{l}{4a }p_1R'(x)
-\frac{l}{2a }p_2W_x(x,y)
-a ^2\frac{\bar{W}(x,y)}{(x-y)^2}+\frac{l^2R''(x)}{12a ^2}
+\frac{\lambda l^2}{3a ^2}
$$
where
$$R(x)=(\lambda_1-\lambda_2)(x^2-1)^2+4(\lambda_3-\lambda_2)x^2,
$$
$$
W(x,y)=\lambda_1 (x^2-1)(y^2-1)-\lambda_2(x^2+1)(y^2+1)+4\lambda_3 x y,$$
$$\bar{W}(x,y)=\lambda_2\lambda_3 (x^2-1)(y^2-1)-
\lambda_3\lambda_1(x^2+1)(y^2+1)+4\lambda_1\lambda_2 xy+\kappa (x-y)^2,$$
$$\lambda=\lambda_1+\lambda_2+\lambda_3,
\qquad \kappa=\lambda_1\lambda_2+\lambda_2\lambda_3+\lambda_3\lambda_1.$$
We define Kowalewski variables as the roots
$s_{1}(q_{1},q_{2})$ and $s_{2}(q_{1},q_{2})$ of the equation
$$
(q_{1}-q_{2})^2\, s^2+ W(q_{1},q_{2})\, s+\bar{W}(q_{1},q_{2})=0.
$$
Associating the $t$-dynamics with $K$, we get equations (\ref{s12g1}), (\ref{s12g2})
for $s_1,s_2$, where
$$
S(\xi)=4(\xi-\lambda_1)(\xi-\lambda_2)(\xi-\lambda_3).
$$
Our general procedure for separation of variables from Section 3 leads to curve (\ref{curve}), where
$$
s_6=0,\qquad l(\xi)=\frac{l^2}{64},\qquad 4\,k(\xi)=a^2\xi^2+(2
e_{1}-a^2\lambda)\xi-e_{2}.
$$
Note that for the Clebsch case we have $\delta=i\frac{l}{4a}.$
One has to replace  $arctanh$ by $arctan$ in the formula (\ref{fullsep}) and after that
the action remains to be real.
\hspace{0.5cm}
\section{Case $S=-1$}
In the case $S(\xi)=-1$  common solutions to the pair (\ref{hamnew}), (\ref{intnew})
satisfy the Gibbons-Tsarev equation (\ref{hydro}). In this case
the curve (\ref{curve}) has the form
$$-4\alpha\delta^2 Y^4+(4e_1+4\xi e_2+4\xi^2\alpha)Y^2+1=0. $$
This curve of genus 1 can be parameterized by the Weierstrass function as
follows:
$$Y=-\frac{1}{16\alpha^2}\frac{y}{x-\beta_3},\qquad \xi=-\frac{e_2}{2\alpha}-
\frac{y}{16}\frac{\delta}{\alpha^2}\left(\frac{1}{x-\beta_1}+\frac{1}{x-\beta_2}-\frac{1}{x-\beta_3}\right),
$$
where $x=\wp(z,g_2,g_3),\quad y=\wp '(z,g_2,g_3),$
$$
y^2=4(x-\beta_1)(x-\beta_2)(x-\beta_3),\quad \beta_3=\frac{8\alpha^2}{3\delta^2}(4\alpha e_1-e_2^2),\;
\beta_{1,2}=-\frac{1}{2}\beta_3\pm\frac{16\alpha^3}{\delta}\sqrt{-\alpha}.
$$

The meromorphic integrals in
$$
dt=\sum\limits_{n=1}^3\omega_1(\xi_n),\qquad
d\tau=\sum\limits_{n=1}^3\omega_2(\xi_n)
$$
can be taken. The result is given by
$$
z_1+z_2+z_3=const,\quad t=\frac{1}{2\sqrt{-\alpha}}\sum\limits_{i=1}^3\log
(\frac{x_i-\beta_2}{\beta_1-x_i}),\quad \tau=-\frac{e_2}{\alpha}t-2\alpha\sum\limits_{i=1}^3
\frac{y_i}{(x_i-\beta_1)(x_i-\beta_2)}
$$
where $x_i=\wp(z_i,g_2,g_3),\;y_i=\wp '(z_i,g_2,g_3).$

\end{document}